# DataChat: Prototyping a Conversational Agent for Dataset Search and Visualization

**Fan, Lizhou**      University of Michigan | lizhouf@umich.edu

**Lafia, Sara**      University of Michigan | slafia@umich.edu

**Li, Lingyao**      University of Michigan | lingyaol@umich.edu

**Yang, Fangyuan**      University of Michigan | yangfy@umich.edu

**Hemphill, Libby**      University of Michigan | libbyh@umich.edu

## ABSTRACT

Data users need relevant context and research expertise to effectively search for and identify relevant datasets. Leading data providers, such as the Inter-university Consortium for Political and Social Research (ICPSR), offer standardized metadata and search tools to support data search. Metadata standards emphasize the machine-readability of data and its documentation. There are opportunities to enhance dataset search by improving users' ability to learn about, and make sense of, information about data. Prior research has shown that context and expertise are two main barriers users face in effectively searching for, evaluating, and deciding whether to reuse data. In this paper, we propose a novel chatbot-based search system, DataChat, that leverages a graph database and a large language model to provide novel ways for users to interact with and search for research data. DataChat complements data archives' and institutional repositories' ongoing efforts to curate, preserve, and share research data for reuse by making it easier for users to explore and learn about available research data.

## INTRODUCTION

Because of the volume, variety of formats, and complexity of connections embedded in the scholarly knowledge linked to research data, it is often hard for researchers and research data management (RDM) units to organize research data for discovery (Gregory et al. 2020; Koesten et al. 2021). Recent RDM guidelines emphasize the importance of discoverability and reusability of research data to promote sharing and transparency of scientific findings (National Science Foundation n.d.; National Institutes of Health 2023). Data archives promote research data findability by assigning a Digital Object Identifier (DOI) to each dataset they distribute (Mooney 2011). While assigning datasets machine-readable identifiers and producing standardized metadata marked up with schema.org tags allows datasets to be harvested and aggregated by large-scale services like Google Dataset Search (Brickley et al. 2019), machine readability does not directly help users determine the reuse potential and relevance of datasets (York 2022).

In this paper, we introduce DataChat, a prototype chatbot for interactive dataset search that leverages a scholarly knowledge graph (SKG) to expand the information available for users to query and access when search for data. We developed and tested DataChat using metadata from the Inter-university Consortium for Political and Social Research (ICPSR). ICPSR provides access to over 11,000 datasets in public social science studies and a bibliography of 100,000 data-related publications that have used ICPSR's data.

ICPSR makes a number of linked resources – including datasets, variables, and publications – available for search and discovery (Levenstein and Lyle 2018); however, links between these resources are not made directly visible to users as they search (Lafia et al. 2022; Fan et al. 2022). Users can currently use third-party aggregators (e.g., Google Dataset Search) or ICPSR's web search system, which is built on a Solr index, to search through study-level metadata, codebooks, variables, and publications (Pienta et al. 2018). Most users initiate their searches through ICPSR's "Find Data" webpage, which provides a search box, a word cloud of popular search term topics, a list of the most downloaded datasets, and other features ("Find Data" 2023). When users search via ICPSR's website, they tend to search directly (e.g., by using a study name), orient while searching (e.g., by looking up subject terms while searching), or take scenic approaches (e.g., by navigating to and comparing multiple study datasets) (Lafia et al. 2023). Datasets and publications are directly accessible when their metadata properties match users' queries; relationships between objects are not directly exposed to users.

By contrast, graph databases and scholarly knowledge graphs (SKGs), organize structured information according to relationships between entity types. SKGs link semantic, directed, and labeled networks of entities (nodes) in academic research by their relations (edges), organizing structured scholarly information from a variety of unstructured sources (Verma et al. 2023; Auer and Kasprzik 2018). To facilitate DataChat, we developed an SKG for ICPSR (ICPSR-SKG) that encodes the same metadata currently available through the ICPSR search system and enables new interactions with research datasets in three main ways. First, the ICPSR-SKG explicitly stores context about the relationships between entity types (e.g., publications and datasets) that users can access, explore, and query. Second, the SKG renders interactive network visualizations, which support user understanding of large-scale relationships across entity types. Finally, unlike systems that are built on static indexes, the SKG is built on top of a graph database, which supports natural language understanding that leverages the connections within the data.

SKGs support applications, like conversational or collaborative chatbots, that work with users to explore and navigate linked, scholarly information (Meloni et al. 2021). To maximize the usability of the information encoded in SKGs, we use a large-language model (LLM) to convert users' natural language questions into Cypher queries (an SQL-inspired query language for graphs), which are expressed in a machine-readable database language. LLMs employ billions of parameters and outperform previous natural language processing models (Shen et al. 2023; Zhao et al. 2023; Fan et al. 2023) and are widely applied to chatbot applications (Yu et al. 2021; Day and Shaw 2021; Harmouche et al. 2020). We leverage a specific LLM, GPT-3.5-turbo (OpenAI n.d.), to help users query the ICPSKG database in the DataChat workflow.

DataChat uses the same underlying metadata currently available in ICPSR's dataset search to contribute novel: (1) **front-end interactions for users** (i.e., natural language queries and network visualizations); and (2) **back-end relationships in databases** (i.e., semantic triples). As a conversational assistant to dataset users and other stakeholders, DataChat traverses ICPSR-SKG as the knowledge base for answering users' dataset-related questions. DataChat then presents the resulting textual and visual representations in an interactive user interface, enabling users to explore relationships between research datasets available from ICPSR.

## DATA AND METHODS

We selected the DataChat technology stack shown in **Figure 1** based on our original design goals of: (1) enhancing metadata *context* by exposing links between entity types; and (2) increasing users' proficiency with the search system, regardless of their level of research *expertise*. The search system is centered around **Datasets**, which have explicit contexts derived from their relationships with other scholarly entities, including research **Publications**.

**(a) ICPSRSKG schema**

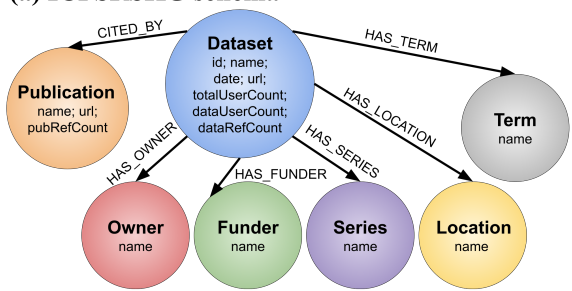

**(b) DataChat workflow**

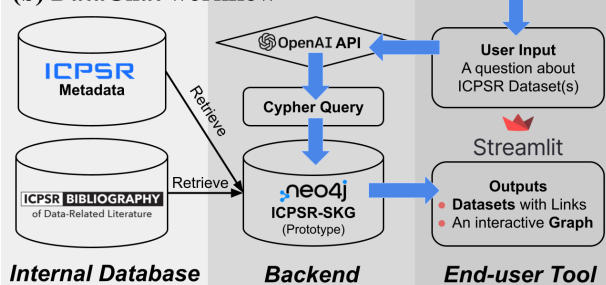

**Figure 1. DataChat design: (a) schema and (b) workflow for the ICPSR-SKG graph database prototype**

As **Figure 1(a)** indicates, the schema of the ICPSR-SKG prototype includes dataset nodes and other types of nodes linked to them. For scalability and experimentation, we selected the 1,642 ICPSR datasets released from 2017 to 2022. Dataset nodes have seven attributes, including the dataset's unique identifier ("id"), its formal study title ("name"), its creation ("date"), the "url" of its DOI, the total number of users who downloaded any metadata or data of the dataset ("totalUserCount"), the number of users who downloaded datasets ("dataUserCount"), and total number of publications that have cited the dataset ("dataRefCount"). The other six types of nodes are linked to dataset nodes through unique types of relations. While all six types of nodes, including publication, owner, funder, series, location, and term, have the "name" attribute, the publication nodes also have the "url" of DOI and the number of citations ("pubRefCount"). We derived publication information from ICPSR Bibliography (ICPSR 2023).

**Figure 1(b)** illustrates the DataChat system design, incorporating a seamless workflow between an end-user tool based on Streamlit (Snowflake Inc. n.d.), a backend processing system utilizing the OpenAI API (OpenAI 2020), and an internal Neo4j-based ICPSR-SKG retrieving data from ICPSR databases (Neo4j, Inc. n.d.). The interaction process starts with the user input, a natural language question about datasets, on the Streamlit interface, which is then sent to the backend for processing using the OpenAI API of the GPT-3.5-turbo model (OpenAI n.d.). The API processes the prompt to generate a Cypher query, the native query language for Neo4j databases, where the prompt is based on the combination of the user input and engineered input-output pairs, e.g., the natural language input "What are the top 5 most cited datasets not owned by ICPSR?" corresponds to the Cypher query output "`MATCH (a:Dataset) WHERE a.owner <> 'ICPSR' RETURN a.name + " LINK: " + a.url AS response ORDER BY a.dataRefCount DESC LIMIT 5`". The ICPSR-SKG Neo4j database then executes the generated Cypher query to retrieve relevant nodes and edges, which are returned to the Streamlit-based interface as either chat messages or a subgraph of the ICPSR-SKG using the streamlit-agraph, a Streamlit Python package that visualizes interactive network graphs (Klose 2023). The **code repository** for the DataChat system is available on https://github.com/casmlab/DataChat.

## RESULTS: THE DATACHAT DASHBOARD

The DataChat dashboard includes two tabs, the DataChatBot Tab and the DataChatViz Tab. Based on users' natural language inputs, these two tabs respectively provide suggestions of datasets with links and visualize interactive graphs for users' exploration. **Figure 2** shows the results of the DataChat dashboard for the example input "What are the latest datasets owned by ICPSR that have been cited by publications more than 3 times?" We intentionally used a grammatically ambiguous query to demonstrate the system's flexibility in query interpretation. The DataChatBot Tab (**Figure 2(a)**) contains three parts, including a question input frame on the bottom, the conversation panel on the top left, and the generated Cypher query on the top right. Users can modify the input and rerun by pressing the "Enter" button. The resulting messages start at the bottom and scroll up, similar to texting, promoting familiarity and ease of use, as most users are already accustomed to this layout. We also keep the Cypher query available to users for transparency, debugging, learning, and feedback purposes.

The DataChatViz Tab (**Figure 2(b)**) is a colored graph visualization where colors correspond to object types (**Figure 1(a)**). In addition to visualizing different node types and names, the graph also highlights the attribute nodes shared by at least two datasets, which are positioned at the center of the graph. For example, the American Health Values Survey and the Massachusetts Health Reform Survey are both owned by HMCA, the Health Management Company of America. Notably, the graph(s) are also interactive – users can highlight and place the nodes and edges for their illustration needs. A **video demo** of the DataChat dashboard is available on https://youtu.be/y4EaJzV2nA8.

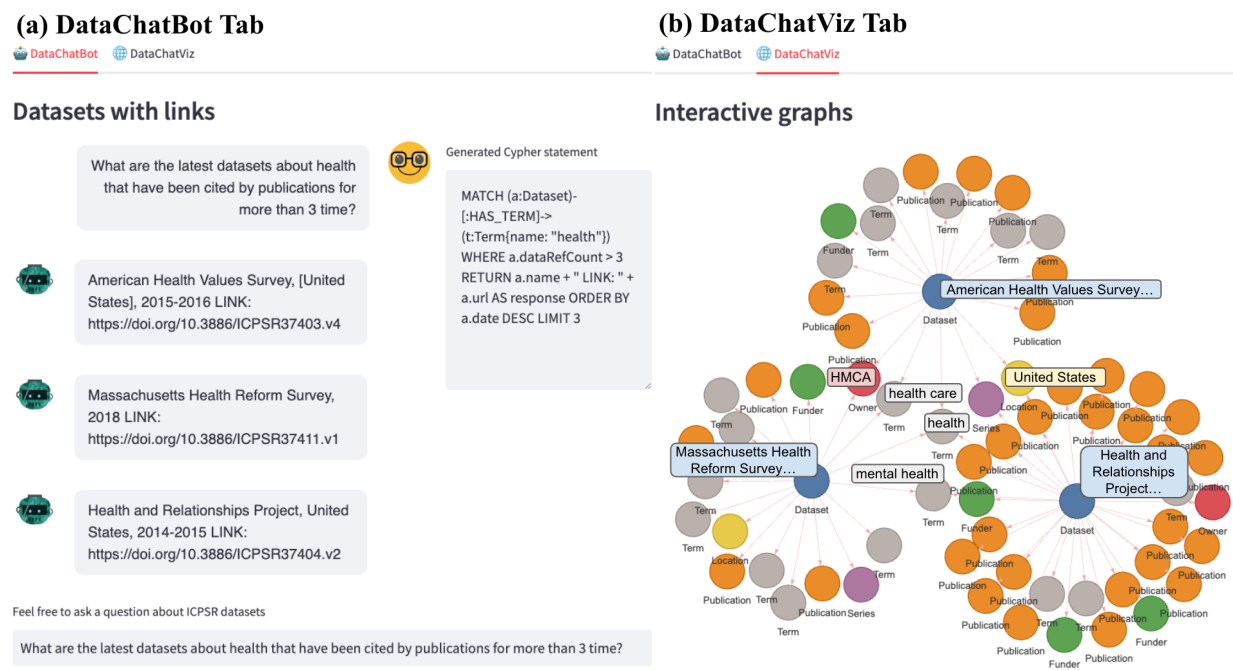

**Figure 2. Prototype DataChat user interface of ICPSR-SKG. Several mouse-over effects in the interactive tab are combined and presented simultaneously, while the actual graph shows them separately.**

## PRELIMINARY EVALUATION

To evaluate the performance of the DataChat prototype, we generated and tested 105 natural language questions about ICPSR datasets. These questions are inspired by a prior study of "genuine information needs" (Papenmeier et al. 2021) for specific social science data stakeholder perspectives from education, funding agencies, and data management units. These questions provide a preliminary evaluation of DataChat's overall ability and versatility. Two authors from our team evaluated the system outputs of the 105 questions and annotated them as "pass" or "not pass". Questions "passed" if they met the expected information need in the original query (semantics of natural language) and compiled a query that can be executed within the existing schema (syntax); they did "not pass" if they returned incomprehensible results or violated the schema. The evaluators reached a high level of agreement in the first round of annotation ($\alpha = 0.87$) (Krippendorff 2018), and reached the full agreement after discussion. As **Table 1** shows, the overall pass rate was 61% (64/105), while the pass rate per stakeholder varied: for education-related (e.g. researchers, educators, and students) queries, the pass rate was 83% which was the highest; for data management-related (e.g. archivists, librarians, and bibliographers) queries, the pass rate was 74%; for funding-related (e.g. governments, NGOs, and corporates) queries, the pass rate was 26%, which was the lowest.

| Pass rate per stakeholder | Input example | Corresponding Cypher query output |
|---|---|---|
| *Education:* 29/35 (**83%**) | What are the most popular datasets about mental health? | `MATCH (a:Dataset)-[:HAS_TERM]->(t:Term) WHERE t.name CONTAINS 'mental health' RETURN a.name + " LINK: " + a.url AS response ORDER BY a.dataUserCount DESC LIMIT 3` |
| *Funding agency:* 9/35 (**26%**) | Which datasets have been funded by the National Institutes of Health or Ford Foundation? | `MATCH (a:Dataset)-[:HAS_FUNDER]->(f:Funder) WHERE f.name IN ["National Institutes of Health", "Ford Foundation"] RETURN a.name + " LINK: " + a.url AS response ORDER BY a.date DESC LIMIT 3` |
| *Data management unit:* 26/35 (**74%**) | Which datasets include information from countries in the Middle East, such as Saudi Arabia or Iran? | `MATCH (a:Dataset)-[:HAS_LOCATION]->(l:Location) WHERE l.name CONTAINS 'Saudi Arabia' OR l.name CONTAINS 'Iran' OR l.name CONTAINS 'Middle East' RETURN a.name + " LINK: " + a.url AS response ORDER BY a.date DESC LIMIT 3` |

**Table 1: Evaluation of DataChat with stakeholder-specific examples**

## DISCUSSION AND CONCLUSION

DataChat is a prototype conversational system for exploring dataset knowledge graphs, and is our initial step towards providing users with needed *context* and bridging gaps in stakeholder *expertise* through *intelligent* and *user-friendly* dataset search. DataChat leverages scholarly knowledge graphs (SKGs), large language models (LLMs), and network visualization to create a novel, interactive data search system. DataChat improves four key aspects of research data search: **Connectivity**, **Efficiency**, **Visibility**, and **Interactivity** (**CEVI**). DataChat also provides a comprehensive research and development workflow from knowledge application design to end-user tool development, which is broadly applicable to data and digital curation applications.

### Essential capabilities of SKGs to support dataset search, data reuse, and management: CEVI

DataChat's enhanced **Connectivity** links ICPSR databases, connects natural language input to ICPSR-SKG, and integrates textual and visual information. These features benefit stakeholders (e.g., archivists, librarians, and bibliographers) by facilitating metadata management and dataset discovery (Djebbar and Belalem 2016; Corrall et al. 2013). The improved **Efficiency** of DataChat, which replaces multiple search dropdowns with a single natural language input, makes the dataset search process accessible for researchers, educators, and students, regardless of their technical expertise and time constraints. DataChat increases data **Visibility** through graph visualization, which also highlights different attributes of nodes and the schema of ICPSR-SKG, enabling stakeholders to evaluate research impacts, identify gaps in knowledge, uncover potential collaborators, and gain insights into emerging research trends (Verma et al. 2023; Manghi et al. 2021). Lastly, DataChat visualization's **Interactivity** promotes user engagement by allowing users to emphasize specific nodes according to their needs and goals, creating a personalized experience as stakeholders explore research datasets.

### LLMs bridge the human-database language gap, while performance varies by stakeholder

DataChat leverages GPT-3.5-turbo, one of the Generative Pre-trained Transformer (GPT) family's LLMs developed by OpenAI (Eloundou et al. 2023), known for their versatility in dealing with unseen scenarios or tasks which are essential abilities of artificial general intelligence. In general, LLMs support usability in SKG applications because they bridge the gap between natural language and graph database queries, enabling researchers to operate in network terms without prior knowledge about a specific type of database language. The GPT-3.5-turbo model works well for example inputs from education and data management unit stakeholders' perspectives. However, our evaluation indicated that the Cypher queries generated for stakeholders in the funding agency are not properly querying data from the ICPSR-SKG, possibly because of the complexity and ambiguity of those stakeholders' interests.

### Research outlook

As we develop the DataChat system, we plan to add details about funder-related attributes, fields of research, and topics of publications that cite data. In addition, while the evaluation examples of natural language queries are helpful, they should be more comprehensive. For example, future evaluation will perform user testing to explain why some queries may not result in anticipated search data. Finally, we will enhance the scalability of the visualization interface, which currently supports three to five datasets in focus.

## ACKNOWLEDGMENTS

This material is based upon work supported by the National Science Foundation under grant 2121789.

## REFERENCES

Auer, Sören, and Anna Kasprzik. 2018. *Towards a Knowledge Graph for Science*. Gottfried Wilhelm Leibniz Universität Hannover.

Brickley, Dan, Matthew Burgess, and Natasha Noy. 2019. "Google Dataset Search: Building a Search Engine for Datasets in an Open Web Ecosystem." In *The World Wide Web Conference on - WWW '19*, 1365–75. San Francisco, CA, USA: ACM Press.

Corrall, Sheila, Mary Anne Kennan, and Waseem Afzal. 2013. "Bibliometrics and Research Data Management Services: Emerging Trends in Library Support for Research." *Library Trends* 61 (3): 636–74.

Day, Min-Yuh, and Sheng-Ru Shaw. 2021. "AI Customer Service System with Pre-Trained Language and Response Ranking Models for University Admissions." In *2021 IEEE 22nd International Conference on Information Reuse and Integration for Data Science (IRI)*, 395–401.

Djebbar, Esma Insaf, and Ghalem Belalem. 2016. "Tasks Scheduling and Resource Allocation for High Data Management in Scientific Cloud Computing Environment." In *Mobile, Secure, and Programmable Networking*, 16–27. Springer International Publishing.

Eloundou, Tyna, Sam Manning, Pamela Mishkin, and Daniel Rock. 2023. "GPTs Are GPTs: An Early Look at the Labor Market Impact Potential of Large Language Models." *arXiv [econ.GN]*. arXiv. http://arxiv.org/abs/2303.10130.

Fan, Lizhou, Lingyao Li, Zihui Ma, Sanggyu Lee, Huizi Yu, and Libby Hemphill. 2023. "A Bibliometric Review of Large Language Models Research from 2017 to 2023." *arXiv [cs.DL]*. arXiv. http://arxiv.org/abs/2304.02020.

Fan, Lizhou, Sara Lafia, David Bleckley, Elizabeth Moss, and Andrea Thomer. 2022. "Librarian-in-the-Loop: A Natural Language Processing Paradigm for Detecting Informal Mentions of Research Data in Academic Literature." *arXiv Preprint arXiv*. https://arxiv.org/abs/2203.05112.

"Find Data." n.d. Accessed April 10, 2023. https://www.icpsr.umich.edu/web/pages/ICPSR/index.html.

Gregory, Kathleen, Paul Groth, Andrea Scharnhorst, and Sally Wyatt. 2020. "Lost or Found? Discovering Data Needed for Research." *Harvard Data Science Review*.

Harmouche, Rola, Aidan Lochbihler, Francis Thibault, Gino De Luca, Catherine Proulx, and Jordan L. Hovdebo. 2020. "A Virtual Assistant for Cybersickness Care." In *2020 IEEE 33rd International Symposium on Computer-Based Medical Systems (CBMS)*, 384–87.

ICPSR. 2023. "Data-Related Publications." 2023. https://www.icpsr.umich.edu/web/pages/ICPSR/citations/.

Klose, Christian. 2023. *Streamlit-Agraph* (version 0.0.45). https://github.com/ChrisDelClea/streamlit-agraph.

Koesten, Laura, Kathleen Gregory, Paul Groth, and Elena Simperl. 2021. "Talking Datasets – Understanding Data Sensemaking Behaviours." *International Journal of Human-Computer Studies* 146 (102562): 102562.

Krippendorff, Klaus. 2018. *Content Analysis: An Introduction to Its Methodology*. SAGE Publications.

Lafia, Sara, Lizhou Fan, Andrea Thomer, and Libby Hemphill. 2022. "Subdivisions and Crossroads: Identifying Hidden Community Structures in a Data Archive's Citation Network." *Quantitative Science Studies* 3 (3): 694–714.

Lafia, Sara, A. J. Million, and Libby Hemphill. 2023. "Direct, Orienting, and Scenic Paths: How Users Navigate Search in a Research Data Archive." In *Proceedings of the ACM on Human Information Interaction and Retrieval (CHIIR)*.

Levenstein, Margaret C., and Jared A. Lyle. 2018. "Data: Sharing Is Caring." *Advances in Methods and Practices in Psychological Science* 1 (1): 95–103.

Manghi, Paolo, Andrea Mannocci, Francesco Osborne, Dimitris Sacharidis, Angelo Salatino, and Thanasis Vergoulis. 2021. "New Trends in Scientific Knowledge Graphs and Research Impact Assessment." *Quantitative Science Studies* 2 (4): 1296–1300.

Meloni, Antonello, Simone Angioni, Angelo Salatino, Francesco Osborne, Diego Reforgiato Recupero, and Enrico Motta. 2021. "AIDA-Bot: A Conversational Agent to Explore Scholarly Knowledge Graphs."

Mooney, Hailey. 2011. "Citing Data Sources in the Social Sciences: Do Authors Do It?" *Learned Publishing: Journal of the Association of Learned and Professional Society Publishers* 24 (2): 99–108.

National Institutes of Health. 2023. "2023 NIH Data Management and Sharing Policy." National Institutes of Health. January 25, 2023. https://oir.nih.gov/sourcebook/intramural-program-oversight/intramural-data-sharing/2023-nih-data-management-sharing-policy.

National Science Foundation. n.d. "Open Data at NSF." National Science Foundation. Accessed March 23, 2023. https://www.nsf.gov/data/.

Neo4j, Inc. n.d. "NEO4J GRAPH DATA PLATFORM | Blazing-Fast Graph, Petabyte Scale." Neo4j. Accessed April 10, 2023. https://neo4j.com/.

OpenAI. 2020. "OpenAI API." OpenAI. September 18, 2020. https://openai.com/blog/openai-api.

———. n.d. "GPT-3.5." OpenAI. Accessed April 10, 2023. https://platform.openai.com/docs/models/gpt-3-5.

Papenmeier, Andrea, Thomas Krämer, Tanja Friedrich, Daniel Hienert, and Dagmar Kern. 2021. "Genuine

Information Needs of Social Scientists Looking for Data." *Proceedings of the Association for Information Science and Technology* 58 (1): 292–302.

Pienta, Amy M., Dharma Akmon, Justin Noble, Lynette Hoelter, and Susan Jekielek. 2018. "A Data-Driven Approach to Appraisal and Selection at a Domain Data Repository." *International Journal of Digital Curation* 12 (2). https://doi.org/10.2218/ijdc.v12i2.500.

Shen, Yiqiu, Laura Heacock, Jonathan Elias, Keith D. Hentel, Beatriu Reig, George Shih, and Linda Moy. 2023. "ChatGPT and Other Large Language Models Are Double-Edged Swords." *Radiology*, January, 230163.

Snowflake Inc. n.d. "A Faster Way to Build and Share Data Apps." Streamlit. Accessed April 10, 2023. https://streamlit.io/.

Verma, S., R. Bhatia, S. Harit, and S. Batish. 2023. "Scholarly Knowledge Graphs through Structuring Scholarly Communication: A Review." *COMPLEX & INTELLIGENT SYSTEMS*. https://doi.org/10.1007/s40747-022-00806-6.

York, Jeremy. 2022. "Seeking Equilibrium in Data Reuse: A Study of Knowledge Satisficing." University of Michigan.

Yu, Shi, Yuxin Chen, and Hussain Zaidi. 2021. "AVA: A Financial Service Chatbot Based on Deep Bidirectional Transformers." *Frontiers in Applied Mathematics and Statistics* 7. https://doi.org/10.3389/fams.2021.604842.

Zhao, Wayne Xin, Kun Zhou, Junyi Li, Tianyi Tang, Xiaolei Wang, Yupeng Hou, Yingqian Min, et al. 2023. "A Survey of Large Language Models." *arXiv [cs.CL]*. arXiv. http://arxiv.org/abs/2303.18223.